\documentclass[aps,prc,groupedaddress]{revtex4}

\usepackage{amsfonts}
\usepackage{graphicx}

\begin{document}


\title{Infrared QCD}


\author{Marco Frasca}
\email[]{marcofrasca@mclink.it}
\affiliation{Via Erasmo Gattamelata, 3 \\ 00176 Roma (Italy)}


\date{\today}

\begin{abstract}
We prove that Nambu-Jona-Lasinio model is an exact description of infrared QCD deriving it from QCD Lagrangian. The model we obtain is renormalizable and confining but, taking very small momenta fixes completely all the parameters of the Nambu-Jona-Lasinio model through those of QCD. The choice of the infrared propagator is done consistently with recent numerical results from lattice and Dyson-Schwinger equations for Yang-Mills theory. The model we get coincides, once the ultraviolet contribution is removed, with the one proposed by
Langfeld, Kettner and Reinhardt [Nucl. Phys. A {\bf 608}, 331 (1996)].
\end{abstract}


\maketitle


\section{Introduction}

Studies of low energy strong interaction processes generally meet severe difficulties due to the known impossibility to manage the right theory, i.e. Quantum ChromoDynamics (QCD), in this limit. So, a lot of phenomenological models have been devised to cope with this problem. One of the most used is the Nambu-Jona-Lasinio model \cite{njl1,njl2,kl,hk,rk,bu}. Despite some severe difficulties this model has,
a large body of 
literature developed about as it reproduces most of the observed patterns in nuclear physics. The main ones are that this model is not renormalizable, one has contact interaction, and fails to maintain quarks confined.

Some years ago, Reinhardt, Kettner and Langfeld \cite{rkl} provided a renormalizable Nambu-Jona-Lasinio model once the one-gluon exchange propagator is known. At the time this latter information was lacking but the authors put out a plausible model for this propagator and were able to perform significant computations in nuclear physics. This model, as we will show, implies that a strong similarity exists, at least in the infrared, between QCD and Quantum ElectroDynamics (QED). Indeed, this deep insight is correct as we will see implying a justification for authors' computations and also for a lot of published literature using Nambu-Jona-Lasinio model that represents the right behavior of hadronic matter in the low energy limit.

In order to have a consistent argument, two main points should be addressed. Firstly, one should know if a Green function method does hold also in a strong coupling regime. We have proved this in a series of papers \cite{fra1,fra2,fra3} showing that exists an expansion for nonlinear equations, that holds at small times, implying the Green function of these equations. When a Fourier transform is done, one gets powers of the energy giving a series that holds when energy goes to zero as we need. Secondly, one has to have a proper gluon propagator, 
that is
the right one from Yang-Mills theory.

About the question of the behavior of the gluon propagator in the infrared there is currently a lot of activity. A first understanding about this matter was put forward by Alkofer and von Smekal (a review is \cite{as}) that, introducing a truncation to the Dyson-Schwinger equations for a pure Yang-Mills, obtained a gluon propagator going to zero in the infrared. A older work due to Cornwall \cite{co} has showed a contradictory result with a gluon propagator reaching a constant value as the energy goes to zero. Both scenarios implied a running coupling reaching a fixed point in the infrared. The present situation has reached a different grasp of the situation mostly through lattice computations and numerical solution of Dyson-Schwinger equations. Recent lattice computations proved that indeed the gluon propagator reaches a non-zero constant value in the infrared but the running coupling does not reach any fixed value but rather it goes to zero \cite{cuc,ilg,ste}. Same happens in the numerical solution of the Dyson-Schwinger equations \cite{an,cs}. A running coupling going to zero has been also seen on lattice computations in \cite{bou1} and with an analysis of experimental data \cite{pro1,pro2,pro3}. So, consensus is going to be reached about a gluon propagator going to a non-zero value as momentum goes to zero,
the ghost propagator being the same of that of a free particle,
and a running coupling, also with different definitions, going to zero. This scenario was already well described by Boucaud's group with lattice computations \cite{bou2,bou3,bou4}. This emerging scenario is in complete agreement with the one we have put out recently \cite{fra4,fra5,fra6,fra7,fra8} based on \cite{fra1}. This gives a gluon propagator in a closed form in very good agreement, as we will show, with lattice computations and with the numerical solution of the Dyson-Schwinger equations while the running coupling obtained with Callan-Symanzik equation \cite{fra8} is in agreement with lattice computations \cite{bou1}.

The paper is so structured. In \ref{sec2} we outline our argument with a reformulation of QED. In \ref{sec3} we apply the same argument to QCD obtaining a renormalizable Nambu-Jona-Lasinio model to be applied in the infrared.
In \ref{sec4a} we discuss the question of the ghost field and Gribov copies in the infrared limit where both seem to be excluded.
In \ref{sec4} we show how our choice of the gluon propagator is indeed a consistent one with respect to lattice and numerical results. In \ref{sec5} we obtain the contact interaction proper to the original formulation of the Nambu-Jona-Lasinio model fixing all the parameters of the theory. We give also some discussion of the ground state of QCD with respect to a pure Yang-Mills theory. Finally, in \ref{sec6} conclusions are given.

\section{\label{sec2} Quantum Electrodynamics}

Let us consider a classical action for the interaction between a charged particle and photons
\begin{equation}
\label{eq:S}
   S = \int d^4x \left[\bar\psi(i\gamma\cdot\partial-e\gamma\cdot A - m)\psi
     -\frac{1}{4}(\partial_\mu A_\nu-\partial_\mu A_\nu)(\partial^\mu A^\nu-\partial^\mu A^\nu)
     -\frac{1}{2\alpha}(\partial\cdot A)^2\right].
\end{equation}
This action gives the following equations
\begin{eqnarray}
   (i\gamma\cdot\partial - e\gamma\cdot A-m)\psi &=& 0 \nonumber \\
   \partial_\mu\partial^\mu A^\nu-\left(1-\frac{1}{\alpha}\right)\partial^\nu\partial^\mu A_\mu 
   &=& e\bar\psi\gamma^\nu\psi
\end{eqnarray}
The solution for the electromagnetic field can be obtained immediately through Green function as
\begin{equation}
\label{eq:field}
   A_\mu(x) = e\int d^4y D^{\mu\nu}(x-y)\bar\psi(y)\gamma_\nu\psi(y)
\end{equation}
being
\begin{equation}
   D^{\mu\nu}(k)=\left[\eta^{\mu\nu}-(1-\alpha)\frac{k^\mu k^\nu}{k^2}\right]
   \frac{1}{k^2+i\epsilon}
\end{equation}
the photon propagator. Substitution of eq.(\ref{eq:field}) into action (\ref{eq:S}) finally gives
\begin{equation}
\label{eq:qed}
   S = \int d^4x\left[\bar\psi(x)(i\gamma\cdot\partial-m)\psi(x)-
   \frac{1}{2}e^2\bar\psi(x)\gamma_\mu\psi(x)\int d^4yD^{\mu\nu}(x-y)\bar\psi(y)\gamma_\nu\psi(y)\right]
\end{equation}
where we can see the appearance of a quartic term. At this stage we can do quantum field theory
with both actions and so we are granted that both are renormalizable. But we can also see that for QED, due to the photon being massless, there is not an immediate way to get a contact quartic interaction out of this latter action. Things can be quite different in QCD as we know that this theory in the infrared develops a mass gap. Finally we note that a generic perturbation term with a propagator in agreement with the K\"allen-Lehman form
\begin{equation}
  G(p)=\sum_n\frac{A_n}{p^2-m^2_n+i\epsilon}
\end{equation}
does not spoil renormalizability of the theory being a sum of renormalizable terms. Indeed, it is enough to have
an infrared gluon propagator reaching a non-null value at zero momenta to recover a Nambu-Jona-Lasinio model (e.g. see \cite{tg}). Our aim in the following is to derive such a formulation for QCD that holds in the infrared. As we will show, we have all one needs to reach this aim.

\section{\label{sec3} Quantum Chromodynamics}

At this stage we work out a classical model. The needs for quantization impose the introduction of a Fadeev-Popov ghost
and to fix the question of Gribov copies. Indeed, in the infrared limit that interests us here the ghost field is
proved to decouple from the gluon field and the Gribov copies are harmless. We will discuss all this matter in
sec. \ref{sec4a}.

So, the action for QCD can be written as
\begin{equation}
   S = \int d^4x\left[\sum_q\bar q\left(i\gamma\cdot\partial - g\frac{\lambda^a}{2}\gamma\cdot A^a - m_q\right)q
   -\frac{1}{4}G^a_{\mu\nu}G^{a\mu\nu}-\frac{1}{2\alpha}(\partial\cdot A)^2\right]
\end{equation}
being $q=u,d,\ldots$ the flavor index and
\begin{equation}
   G^a_{\mu\nu}=\partial_\mu A^a_\nu-\partial_\nu A^a_\mu+gf^{abc}A^b_\mu A^c_\nu
\end{equation}
the field tensor. This action produces the equations
\begin{eqnarray}
\label{eq:ym}
    \left(i\gamma\cdot\partial - g\frac{\lambda^a}{2}\gamma\cdot A^a-m_q\right)q = 0 & &\nonumber \\
    \partial^\mu\partial_\mu A^a_\nu-\left(1-\frac{1}{\alpha}\right)\partial_\nu(\partial^\mu A^a_\mu)
    +gf^{abc}A^{b\mu}(\partial_\mu A^c_\nu-\partial_\nu A^c_\mu)+gf^{abc}\partial^\mu(A^b_\mu A^c_\nu)& & \nonumber \\
    +g^2f^{abc}f^{cde}A^{b\mu}A^d_\mu A^e_\nu = g\sum_q\bar q\frac{\lambda^a}{2}\gamma_\nu q.& &
\end{eqnarray}
At this stage we use the small time approximation devised in \cite{fra1,fra2,fra3} and write down the solution for the field equations as
\begin{equation}
    A^a_\mu(x)\approx g\int d^4y  D_{\mu\nu}^{ab}(x-y)\sum_q\bar q(y)\frac{\lambda^a}{2}\gamma^\nu q(y)
\end{equation}
while for the gluon propagator we take the form
\begin{equation}
   D_{\mu\nu}^{ab}(k)=\delta^{ab}\left[\eta_{\mu\nu}-(1-\alpha)\frac{k_\mu k_\nu}{k^2}\right]G(k).
\end{equation}
This approximation implies that we get a theory that holds at small momenta. Indeed, this is just the leading order of a small time series \cite{fra3}. Then we can write the QCD action that holds in the infrared limit as
\begin{equation}
\label{eq:qcdref}
   S \approx \int d^4x\left[\sum_q \bar q(x)(i\gamma\cdot\partial-m_q)q(x)
   -\frac{1}{2}g^2\sum_{q,q'}\bar q(x)\frac{\lambda^a}{2}\gamma^\mu q(x)
   \int d^4y D_{\mu\nu}^{ab}(x-y)\bar q'(y)\frac{\lambda^b}{2}\gamma^\nu q'(y)\right].
\end{equation}
We just note that Nambu-Jona-Lasinio model is recovered if the gluon propagator has a mass gap $m$. In this case one should have e.g.
\begin{equation}
   G(k)=\frac{1}{k^2-m^2+i\epsilon}
\end{equation}
being $m$ the mass gap. Then, in the infrared $m\gg k$ and then $G(k)\approx constant$ giving back a propagator proportional to $\delta^4(x-y)$ producing the required contact interaction while the mass gap becomes a natural cut-off for the theory. As already said, for QCD this is indeed the situation. 

Then, our next aim is to fix the form of $G(k)$ in agreement with the current scenario in lattice computations for a pure Yang-Mills theory.

\section{\label{sec4a} Ghost field and Gribov copies}

In order to have an idea on how to fix the gauge in the infrared limit, we need to know some results that have been obtained recently
on the lattice. It should be said that, in some cases, analytical proofs exist but we will avoid these as theoretical results need
some time before their correctness is properly assessed. So, we heavily rely on lattice computations that presently, with the use
of huge volumes as we will see in sec.\ref{sec4}, have definitely set the scenario about the propagators at lower momenta.

The question of the importance of Gribov copies in this limit has been present since the start of studies of QCD infrared limit.
Gribov copies are not important in the ultraviolet limit as otherwise we have had serious problems to unveil asymptotic
freedom and compute higher order corrections, a well-acquired technology these days \cite{qcd08}. In the infrared limit there has
been some theoretical work to understand confinement and this set the agenda for people working on the lattice as people
have had to understand the relevance of Gribov scenario in this framework.

The most important works on this matter have been published by Ilgenfritz, M\"uller-Preussker and collaborators \cite{ilg1,ilg2}
quite recently where they consider different volumes and $\beta$ to understand the effect of Gribov copies on gluon and
ghost propagators. Their conclusions are somewhat unexpected: They do not see any effect of Gribov copies on the gluon
propagator but the effect appears quite significant for the ghost propagator. From their later paper \cite{ilg2} they
show as the effect of the Gribov copies is even smaller as the volume enlarges. Indeed, eigenvalues of the
Fadeev-Popov operator are seen accumulating to zero while the ghost propagator saturates at smaller volumes where the
effect of Gribov copies seems more effective. So, the question to be understood after these analyses is how relevant
is the ghost propagator in the behavior of Yang-Mills theory in the infrared.

The answer appeared quite recently in a pair of beautiful papers due to Cucchieri and Mendes \cite{cuc1,cuc2}. They firstly
showed that the gluon propagator must reach a finite non-null value in the infrared limit \cite{cuc1}. Then, they proved
that the ghost propagator behaves as that of a free particle \cite{cuc2}, that means that it decouples from the gluon field and we can avoid
to care about it in the infrared limit. This means that Gribov copies are not relevant at lower momenta in the same
way as happens in the ultraviolet limit. So, we conclude that our starting Lagrangian is well supported by
lattice computations and no further constraints are needed to fix it.

\section{\label{sec4} Gluon propagator}

The choice of the gluon propagator must be consistent with recent lattice computations \cite{cuc,ilg,ste} and in perfect agreement with the numerical solution of the Dyson-Schwinger equations \cite{an}. For our aims we do not consider the corresponding numerical solution on a torus \cite{cs}. We just note that in this case a strongly infrared depressed running coupling is observed. The next most important point is that the gluon propagator must satisfy Callan-Symanzik equations and the corresponding beta function should give a running coupling going to zero in the infrared as is seen on lattice. Finally, we do not have to consider coupling with the ghost as this is seen to behave as a free particle and so decouples from the gluon field.

We have proved in a series of a paper \cite{fra4,fra5,fra6,fra7,fra8} that a pure Yang-Mills theory admits a gluon propagator given by
\begin{equation}
\label{eq:prop}
    G(k)=\sum_{n=0}^\infty\frac{B_n}{k^2-m_n^2+i\epsilon}
\end{equation}
being
\begin{equation}
    B_n=(2n+1)\frac{\pi^2}{K^2(i)}\frac{(-1)^{n+1}e^{-(n+\frac{1}{2})\pi}}{1+e^{-(2n+1)\pi}},
\end{equation}
and
\begin{equation}
\label{eq:ms}
    m_n = \left(n+\frac{1}{2}\right)\frac{\pi}{K(i)}\left(\frac{Ng^2}{2}\right)^{\frac{1}{4}}\mu
\end{equation}
being N the number of colors, $\mu$ an integration constant to be fixed experimentally and $K(i)$ the constant
\begin{equation}
    K(i)=\int_0^{\frac{\pi}{2}}\frac{d\theta}{\sqrt{1+\sin^2\theta}}\approx 1.3111028777.
\end{equation}
This propagator is the same of a massless scalar field theory obtained in the same limit in \cite{fra1}. They coincide as the only useful solution of a Yang-Mills propagator to build a quantum field theory is the one having all equal components \cite{fra5}.  
The reason for this mapping between a massless scalar field and Yang-Mills theory in the infrared limit relies on the fact proved
in the eighties that most solutions of Yang-Mills quantum mechanics are generally chaotic \cite{sav1,sav2,sav3} and so, completely useless to build
a quantum field theory out them. So, these prove to be the only classical solutions of Yang-Mills theory to be used in the infrared limit.
Indeed, these solutions are exact solutions
of classical Yang-Mills equations producing a mass gap out of massless equations of motion \cite{fra9}.
In this case the Yang-Mills theory coincide with the massless scalar field theory provided one substitutes the coupling $\lambda$ with the 't Hooft coupling $Ng^2$.

This propagator fulfills all the requirements  
seen on lattice computations. It reaches a finite non-null value at lower momenta and
recovers the ultraviolet behavior being $\sum_{n=0}^\infty B_n=-1$.
It should also be  noted that
\begin{equation}
\label{eq:jnlc}
   G(0)=\frac{1}{\sigma}\sum_{n=0}^\infty\frac{4}{2n+1}
   \frac{(-1)^{n}e^{-(n+\frac{1}{2})\pi}}{1+e^{-(2n+1)\pi}}\approx 3.761402959\cdot\frac{1}{\sigma}.
\end{equation}
where we have introduced the string tension as $\sqrt{\sigma}=\left(Ng^2\mu^4/2\right)^{\frac{1}{4}}$.
Eq.(\ref{eq:jnlc}) introduces a fundamental constant that defines the Nambu-Jona-Lasinio model from QCD.

In order to give a complete proof of the fact that this is indeed a proper choice for the gluon propagator, we want to compare it with lattice results and the numerical solution of the Dyson-Schwinger equation. While in the former case we expect some error due to volume effects in the lattice computations that should appear to decrease with increasing volume, for the numerical solution of the Dyson-Scwhinger equation we must have a complete coincidence. We will show that is indeed what happens. We emphasize that in order to do this comparison we only need to fix the parameter ${\mu}$ that gives the gluon mass. But we are comparing non homogeneous data belonging to different research groups so the value at zero momenta can be different. This should be expected but the only relevant thing is that fixing just the value of our parameter we are able to get excellent agreement with lattice and numerical data.

\begin{figure}[tbp]
\begin{center}
\includegraphics[width=320pt]{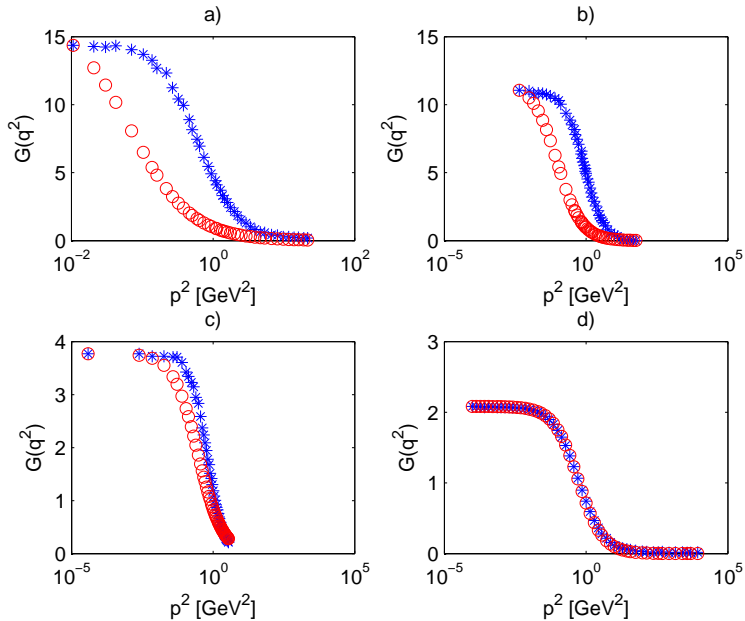}
\caption{\label{fig:fig1} Gluon propagator (circles) compared to lattice data (stars): a) \cite{ilg} $(13.2fm)^4$;
b) \cite{ste} $(19.2fm)^4$; c) \cite{cuc} $(27fm)^4$; d) \cite{an} Dyson-Schwinger.}
\end{center}
\end{figure}

For lattice computations we have that in \cite{cuc} the gluon propagator is computed with the largest volume $(27fm)^4$, the next one is \cite{ste} $(19.2fm)^4$ and finally \cite{ilg} $(13.2fm)^4$. For the numerical solution of the Dyson-Schwinger equations \cite{an} the curves practically coincide. But these computations have not all the same $\sigma$ being different groups. What we want to see is that, independently on the value of $\sigma$, in any case the fits improve
increasing the volume.
This is exactly what happens as can be seen in fig. \ref{fig:fig1} where comparisons with all propagators is done. 

\begin{figure}[tbp]
\begin{center}
\includegraphics[width=320pt]{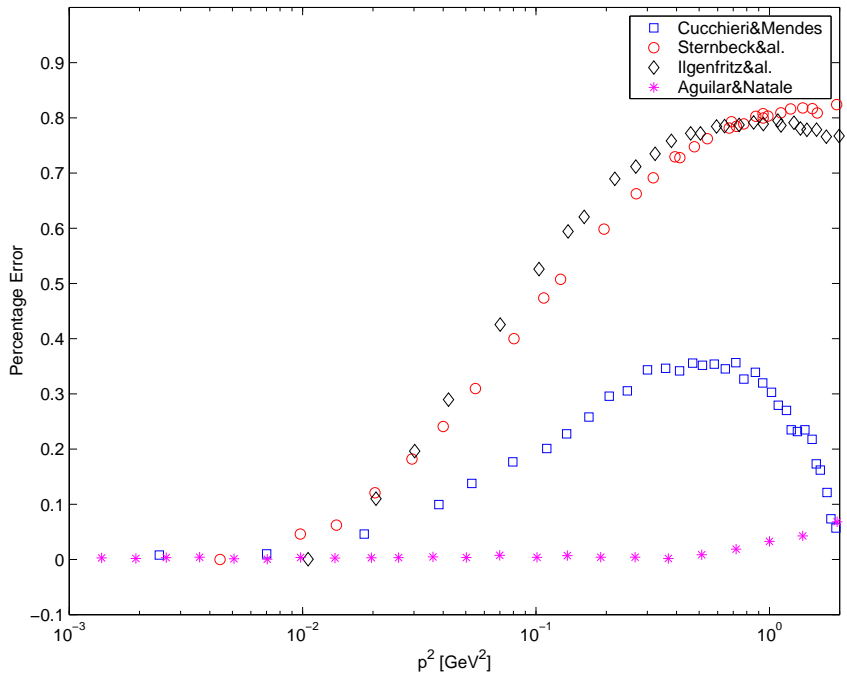}
\caption{\label{fig:fig2} Percentage error for our propagator with respect to lattice and numerical data. It is seen how increasing volume makes converging the data toward our solution.}
\end{center}
\end{figure}

Indeed, one can see from fig. \ref{fig:fig2} that the error is significant in the intermediate energy region and not for infrared but this error significantly decreases going from the smallest to the largest volume reaching a fairly good agreement with the results given in \cite{cuc} and a perfect agreement with numerical Dyson-Schwinger equations \cite{an}. The trend is that the lattice curve indeed converges toward the propagator we have given here at increasing volume.

Then, we can assume that the propagator (\ref{eq:prop}) represents correctly the behavior of Yang-Mills theory in the infrared and we can take this into our rewriting of QCD given in the action (\ref{eq:qcdref}). As this propagator is just a sum of massive free propagators, we are able to get a Nambu-Jona-Lasinio model describing low energy QCD.

\section{\label{sec5} Nambu-Jona-Lasinio model from QCD}

At this stage we can write down the action of a renormalizable Nambu-Jona-Lasinio model as 
\begin{equation}
\label{eq:qcdnjl}
   S_{NJL} \approx \int d^4x\left[\sum_q \bar q(x)(i\gamma\cdot\partial-m_q)q(x)
   -\frac{1}{2}g^2\sum_{q,q'}\bar q(x)\frac{\lambda^a}{2}\gamma^\mu q(x)
   \int d^4yG(x-y)\bar q'(y)\frac{\lambda^a}{2}\gamma_\mu q'(y)\right].
\end{equation}
where use has been made of current conservation that grants independence on the choice of the gauge in the propagator. Here $G(x-y)$ is just the Fourier transform of the propagator in eq.(\ref{eq:prop}).  We emphasize that this model should holds just at lower momenta. This situation is quite different from the case of QED.

At this stage we just notice that our propagator (\ref{eq:prop}) implies the exchange of particles with masses given in eq.(\ref{eq:ms}) and this are not gluons but glueballs that are not confined but would be stable particles if the ground state of QCD would be higher than the lowest of the masses of these particles. This is not the case in the real world and glueballs decay. The lowest glueball state has an energy higher than the $\pi$ meson mass. Then, $\pi$ mesons should appear as decay products of glueballs. With the Nambu-Jona-Lasinio model we should be able to compute such decay rates.

We point out that model given in eq.(\ref{eq:qcdnjl}) is renormalizable due to its strict analogy with QED (\ref{eq:qed}), 
the only difference being a sum of massive propagators that in the limit of high momenta reduces to a sum of QED Lagrangians.
But we can also conjecture that this model is confining due to the nature of Yukawa propagators involved into the interaction terms.
These properties are not maintained with a contact interaction. Indeed, this can be recovered by taking the limit of small momenta (infrared limit $k\rightarrow 0$) giving after a Fourier transform
\begin{equation}
   G(x-y)\approx 3.76\cdot\frac{1}{\sigma}\delta^4(x-y)
\end{equation}
that gives back the original Nambu-Jona-Lasinio model
\begin{equation}
\label{eq:orinjl}
   S \approx \int d^4x\left[\sum_q \bar q(x)(i\gamma\cdot\partial-m_q)q(x)
   -\frac{1}{2}G_{NJL}\sum_{q,q'}\bar q(x)\frac{\lambda^a}{2}\gamma^\mu q(x)
   \bar q'(x)\frac{\lambda^a}{2}\gamma_\mu q'(x)\right].
\end{equation}
being
\begin{equation}
    G_{NJL}\approx 3.76\cdot\frac{g^2}{\sigma}
\end{equation}
the Nambu-Jona-Lasinio constant linked to the experimental value of the string tension $\sigma$. We can see a strong similarity with the electro-weak part of the standard model that in the low energy limit recovers the Fermi model. Here we get again a non-renormalizable model and we do not have to worry about the missing of confinement as at increasing momentum we have to use the more general model given in eq.(\ref{eq:qcdnjl}) and correct it when we are at the boundary of the intermediate energy region. Besides, the model has a natural cut-off originating from the mass spectrum of the Yang-Mills theory. Indeed we have assumed $\frac{\pi}{2K(i)}\sqrt{\sigma}\gg k$ and we can take for the Nambu-Jona-Lasinio cut-off
\begin{equation}
    \Lambda_{NJL}=\frac{\pi}{2K(i)}\sqrt{\sigma}
\end{equation}
that is about the mass of the $\sigma$ particle (or f0(600))  (about 527 MeV) when we take for $\sqrt{\sigma}$ the value 440 MeV currently chosen on lattice computations.


We emphasize that when $|k|>\Lambda_{NJL}$, our approximation fails and higher order corrections are needed.

As a final consideration we note that this is exactly the model presented in Ref.\cite{rkl} when the ultraviolet part is removed. Then, we can see that these authors hit the right model and all their conclusions apply provided we consider a momentum $|k|<\Lambda_{NJL}$.

\section{\label{sec6} Conclusions}

We have shown how, from QCD, a renormalizable Nambu-Jona-Lasinio can be derived. This is possible as, in the infrared limit, we can still adopt the concept of a Green function. The existence of a mass gap that changes the propagator into a massive one, or a sum of free massive propagators, grants the emerging of a contact interaction in the small momenta limit.

It is really surprising the insight of Nambu and Jona-Lasinio that about fifty years ago were able to postulate a model that turns out to be the right one to describe strong interactions in the lower energy limit. A large body of literature has been published about and the main conclusion is that this research activity has been well deserved. On the same ground one should put the more recent work by Langfeld, Kettner and Reinhardt that postulated a renormalizable version of the Nambu-Jona-Lasinio model turning out exactly the one we got from QCD.

With this formulation of low energy QCD, the possibility is given to perform all kind of computations one needs to understand nuclear matter. The model we obtained relies on some experimental values but these are exactly the same needed for QCD as it should be.


\end{document}